\documentclass[aps,prl,preprint,groupedaddress,showpacs]{revtex4}
\usepackage{amssymb}

\begin{document}

\title{Surface plasmon polaritons assisted diffraction in periodic
subwavelength holes of metal films with reduced interplane coupling}
\author{Xu Fang, Zhiyuan Li, Yongbing Long, Hongxiang Wei, Rongjuan Liu,
Jiyun Ma, M. Kamran, Huaying Zhao, Xiufeng Han, Bairu Zhao, and
Xianggang Qiu }
 \email{xgqiu@aphy.iphy.ac.cn}
\affiliation{Institute of Physics, Chinese Academy of Sciences, and Beijing National Laboratory
for Condensed Matter Physics, Beijing 100080, China}
\date{\today}

\begin{abstract}
Metal films grown on Si wafer perforated with a periodic array of
subwavelength holes have been fabricated and anomalous enhanced transmission
in the mid-infrared regime has been observed. High order transmission peaks
up to Si(2,2) are clearly revealed due to the large dielectric constant
contrast of the dielectrics at the opposite interfaces. Si(1,1) peak splits
at oblique incidence both in TE and TM polarization, which confirms that
anomalous enhanced transmission is a surface plasmon polaritons (SPPs)
assisted diffraction phenomenon. Theoretical transmission spectra agree
excellently with the experimental results and confirm the role of SPPs
diffraction by the lattice.
\end{abstract}

\maketitle

Since the discovery of anomalous enhanced transmission of subwavelength
metallic hole arrays in 1998 \cite{NAT391P667}, enormous interest has been
sparked among world wide researchers due to its potential application in
subwavelength optoelectronic devices as well as the underlying physics.
Although the exact origin of the enhanced transmission remains under
controversial \cite{PRL96P213901,PRB66P195105}, it is generally acknowledged
that surface plasmon polaritons (SPPs) are playing a crucial role. Recent
researches indicate that this phenomenon could be a SPPs assisted
diffraction process \cite{PRL92P107401,JOA8P458}. In this process, the
coupling of SPPs on the opposite interfaces of the metal film helps the
electromagnetic energy to tunnel through the subwavelength holes \cite%
{PRL86P1114}. This interplane coupling is substantial when the dielectrics
at the interfaces are similar, and is the strongest when they are identical
\cite{OpC200P1}. To achieve high transmittance, most previous work adopts
either free standing films or films on substrates with small dielectric
constants. However, the strong interplane interference in this kind of
samples will hinder the mechanism from getting fully revealed \cite%
{JOSAA22P998}. For thin films with identical interfaces, this interference
could even lift the degeneracy of SPPs on the two interfaces and alter their
modes \cite{PRL47P1927,PRB72P075405}. So it is of great interest to reduce
the interference in order to study the underlying physics of the SPPs at
each individual interface. For this purpose, we adopt the Air/Au/Si system
in which air and Si have large difference in their dielectric constants in
the mid-infrared. The thickness of the metal films is an order larger than
the skin depth (about 20 nm for Au at the mid-infrared). It is expected that
in such system the interference of the SPP modes at two interfaces is
greatly reduced and the characteristics of SPPs at a single interface are
preserved.

Metal films were deposited on Si wafer by magnetic sputtering. The hole
arrays were fabricated by conventional ultraviolet photolithography and
reactive ion etching. All the patterns are square arrays of circular holes,
occupying an area of 7$\times$7 mm$^{2}$ respectively. The hole number for
each pattern is over one million and the finite size effect is assumed to be
well avoided \cite{APL84P2742}. The infrared transmittance spectra were
obtained with an ABB Bomem DA8 Fourier transform infrared spectrometer.

Shown in Fig. 1(a) is the zeroth-order transmittivity (transmittance
normalized by the porosity of the film) of three Au films under normal
incidence. All three films have a thickness of 320 nm and a hole diameter of
3 $\mu $m. The lattice constants are 5 (sample A), 6 (sample B) and 7
(sample C) $\mu $m respectively. Three salient features
can clearly be discerned. (i) Because the dielectric constant of Si ($%
\varepsilon $=11.7) is much larger than that of air in the mid-infrared
regime, the distance between Air(1,0) and Si(1,0) is greatly expanded in
comparison with previous reports (Si(1,0) are at 21.02 $\mu $m and 24.41 $%
\mu $m for samples B and C respectively, which are not shown in the figure).
(ii) High order peaks such as Si(1,1), (2,0) and (2,1) can be clearly
observed between Air(1,0) and Si(1,0). (iii) Peaks with the same order of
the three samples shift according to the periodicities, and they have larger
wavelengths for larger periodicities. Numerical calculations on the
transmission spectra have been performed by means of a transfer-matrix
method combined with an analytical modal solution \cite{JAP97P033102}, using
the same material parameters as in the experiments, and the results are
displayed in Fig. 1(b). The circular holes are treated as square ones with
the same area, considering the convenience in modeling. A series of resonant
peaks can be clearly found and their positions agree well with the
experimental data.

We attribute the above observation to be a result of SPPs assisted
diffraction process. SPPs have different momenta from light of the
same energy \cite{Raether}. The gap between the momenta can be
compensated with the presence of a periodic hole array. The momentum
matching condition is
\begin{equation}
\vec{k}_{sp}=\vec{k}_{\parallel }\pm i\vec{G}_{x}\pm j\vec{G}_{y},
\label{eq1}
\end{equation}%
where $k_{sp}$ is the wave vector of SPPs, $k_{\parallel }$ is the inplane
wave vector component of the incident light (zero for normal incidence), i
and j are integers, $G_{x}$ and $G_{y}$ are the reciprocal vectors \cite%
{PRB58P6779}. These two indices have been used to designate the resonant
peaks in Fig. 1(a). We have extracted the experimental SPPs dispersion from
the data shown in Fig. 1(a) through Eq. (1) and compared with the
theoretical one which adopts the material parameters in Ref. \cite%
{Palik,ApO22P1099}, and found that these two curves fit very well.

To further confirm the view above, we measured the angular dependent
transmittivity of sample D (thickness 220nm, periodicity 5 $\mu $m and hole
diameter 3 $\mu $m). The incident angle varies from 0$^{\circ }$ to 36$%
^{\circ }$, with a 4$^{\circ }$ increment. Under TM polarization
(polarization perpendicular to the rotation pole), Si(1,0) and Si(1,1) split
into two branches. The data for Si(1,1) is shown in Fig 3. Under TE
polarization (polarization parallel to the rotation pole), Si(1,0) undergoes
minor blueshift (from 17.63 $\mu $m at 0$^{\circ }$ to 17.46 $\mu $m at 36$%
^{\circ }$) while Si(1,1) splits with the splitting getting larger with the
increasing angle, as shown in Fig 2(a). This is the only report that shows
peak splitting under TE polarization up to now. The theoretical spectra, as
displayed in Fig. 2(b), reproduce the experimental data excellently.

SPPs have p-wave like character and the electric field lies parallel
with their propagation direction. They can be excited if the
incident light has a
component of the electric field parallel to their propagation direction \cite%
{PRL86P1110}. For the fourfold degenerate ($\pm$1,0) and (0,$\pm$1) SPP
modes, only (0,$\pm$1) modes can be excited under TE-polarized incident
light. $k_{\parallel}$ has no projection along the directions (0,$\pm$1), so
the modes remain degenerated and no splitting can be observed upon the
change of incident angle. This is what we observed above and by other groups
\cite{PRL92P107401}. However, for Si(1,1) modes which propagate in the
direction ($\pm$1, $\pm$1), $k_{\parallel}$ and $E_{\parallel}$ have a 45$%
^{\circ}$ or 135$^{\circ}$ angle with respect to the propagation direction
of SPPs respectively. All 4 modes can be excited and $k_{\parallel}$ has a
component along their propagation directions. According to Eq. (1), $%
k_{\parallel}$ will be added to or subtracted from the reciprocal vectors,
depending on the propagation direction. So the degeneracy of the ($\pm$1,$%
\pm $1) modes will be lifted off, and the Si (1,1) peak shows a splitting.
The dispersion curves for both polarizations are displayed in Fig. 3. The
theoretical ones extracted from Fig. 2(b) and calculated based on SPPs model
are also presented. The experimental curves for TE and TM polarization are
similar and fit fairly well with the theoretical ones. It is noticed that,
the two peaks under TM polarization are broader and weaker than their
counterparts under TE polarization. This is because TM light penetrates
deeper into the metal and gets absorbed more strongly than TE light.

The similar splitting characters of Si(1,1) under TE and TM
polarizations clearly confirm the involvement of SPPs during the
transmission process. We notice that Ref. \cite{JOSAA22P998} has
successfully predicted this experiment results through Huygens
diffraction model. According to Refs. \cite{JOSAA22P998,OpE12P3652},
we believe that strong interplane coupling would hinder the observation of the splitting behavior
under TE polarization. The strong coupling could be the reason why
in the previous work no splitting in the TE polarization has been
reported.

The discussion above also applies well to the high order peaks of Si(2,0),
(2,1) and (2,2). Under TE polarization, Si(2,0) makes minor blueshift (from
8.66 $\mu$m at 0$^{\circ}$ to 8.47 $\mu$m at 36$^{\circ}$), which is similar
to Si(1,0). Si(2,1) stays nearly unchanged since it will split into 4
branches, making the splitting hard to detect. Si(2,2) is at 6.35 $\mu$m at 0%
$^{\circ}$, then splits into two branches, which are at 5.95 $\mu$m and 6.63
$\mu$m at 16$^{\circ}$. This behavior is similar to Si(1,1). The change for
high order peaks is generally not as significant as the lower order ones.
The reason is that the reciprocal vectors for high order peaks are large,
making them less sensitive to the change of the inplane vector of the
incident light.

To verify that the interplane coupling is indeed very weak in the
present system, we fabricated a series of Pt films with different thickness and did the same measurements. 
All the peaks stay unchanged in wavelength until being overlapped by other peaks or merged into the background as the thickness decreases.
Further numerical calculations on sample D with
various thicknesses are made. The calculated transmission spectra of
these samples under TE-polarized illumination at an incident angle
of 16$^{\circ }$ are displayed in Fig. 4. The resonant peaks stay
unchanged in wavelength while the thickness changes over 7 times
from 120 nm to 920 nm, which clearly manifests the character of
decoupled SPPs \cite{APL81P4327,PRL83P2845}. The splitting behavior
does not change with the thickness, as long as the interplane
coupling is weak. 

We notice that although most existing relevant theories treat the holes in
the array as point scattering centers, it is expected that the finite size
of the hole diameter will modify the exact behavior of the diffraction \cite%
{SCI305P847,PRL94P110501,PRL92P183901,PRL95P103901}. It is found that when
the diameter of the holes is larger than half of the periodicity, the (1,0)
peak will split into two \cite{APL88P213112}. The reduced interference
between the SPPs at two interfaces in the current system facilitates the
study of the influence of hole diameter on the SPPs assisted diffraction
process. To this end, we fabricated sample E with thickness of 320 nm,
lattice constant of 6 $\mu$m, and hole diameter of 4 $\mu$m, and compare the
zeroth-order normal-incidence transmission of sample B and E, which are only
different in hole diameters, as shown in Fig. 5. The most distinguished
difference between the two spectra is around 10 $\mu$m where Si(2,0) and
(2,1) in sample B seem to get merged in sample E. The theoretical result
confirms such a phenomenon, as can be found in the inset.

Since SPPs on the opposite interfaces overlap greater for larger
holes, one may attribute this phenomenon to the enhancement of the
interplane coupling. We believe it is not the real reason. First,
the wave vectors of SPPs corresponding to Si(2,0) and (2,1) are
quite different from those of Air(1,0). There could not be any
strong coupling. Second, we have done the angular dependence
experiments of sample D and obtain similar
results. As discussed previously, this verifies that interplane
coupling is weak in these samples.

We attribute this change to the inplane Bragg scattering of SPPs. When the
hole diameter exceeds half of the periodicity, Bragg scattering will
introduce band gaps to the dispersion relation, which correspond to
different distributions of charges and electromagnetic field in space \cite%
{PRB67P035424}. SPPs corresponding to the high order peaks have small
wavelengths, so they are more sensitive to the actual size of the holes. We
also find through numerical calculation that high order peaks are usually
sharper and more pronounced when the hole diameter is smaller.

In summary, by adopting Si which has a large dielectric constant in
comparison to air in the mid-infrared regime and metal films of
large thicknesses, we succeed in investigating the properties of
SPPs at individual interfaces. The enhanced transmission is observed
in an extended wavelength regime and transmission peaks with highest
order up to (2,2) are clearly revealed and fit well with the inplane
propagation characters of SPPs. The splitting of peaks Si(1,1) is
observed in TE and TM polarization, which confirms that the enhanced
transmission is a SPPs assisted diffraction phenomenon. Numerical
calculations on the transmission spectra have been made and the
results agree excellently with the experimental data. The agreement
between the numerical calculations and experimental results, together with the successful explanation of the experimental results based on SPPs further
clearly confirms the role played by SPPS in the enhanced transmission. Since Si is the base material for modern microelectronic
applications, the extended working wavelength regime for the
subwavelength optics makes the current system in study a potential
prototype element in the future plasmonic circuits which can merge
photonics and electronics at nanoscale dimensions.

\begin{acknowledgments}
The authors thank M. P. van Exter, C. Genet, W. M. Liu and J. J. Hu for
enlightening advice and C. Z. Bi, Q. Luo, and H. F. Yang for technical help.
This work is supported by National Science Foundation of China (No. 10674168
and 10525419) and the MOST of China(973 project No:2006CB601006), and project ITSNEM of the Chinese Academy of Sciences.
\end{acknowledgments}

\newpage

Figure Captions:

Fig. 1 (a) Experimental and (b) theoretical results of
transmittivity of subwavelength hole arrays on Au films. Air stands for
the Air/Au interface and Si stands for the Si/Au interface. The
periodicities are 5 (sample A), 6 (sample B) and 7 (sample C)
$\protect\mu$m respectively. The resonant peaks have been designated
according to the SPPs model.

Fig. 2 (a) Experimental and (b) theoretical results of
transmittivity of sample D under TE-polarized illumination. Shown in
the figure are representative spectra at incident angles from
0$^{\circ}$ to 32$^{\circ}$ with a step of 8$^{\circ}$. Consecutive
curves have been shifted a value of 0.04 upwards to improve
readability.

Fig. 3 (Color online) Dispersion relation of Si(1,1) of sample D.
For TE and TM polarizations, the incident angle in the experiment changes from 0$%
^{\circ}$ to 36$^{\circ}$ with a step of 4$^{\circ}$. The numerical
calculation result is extracted from Fig. 2(b). The line is the
theoretical curve calculated by SPPs model.

Fig. 4 (Color online) Theoretical transmittivity of sample D with
various film thickness. The illumination is TE-polarized, with an
incident angle of 16$^{\circ}$.

Fig. 5 (Color online) Experimental and theoretical (inset) results
of transmittivity of sample B (diameter= 3 $\protect\mu$m) and E (diameter= 4 $%
\protect\mu$m) under normal illumination.

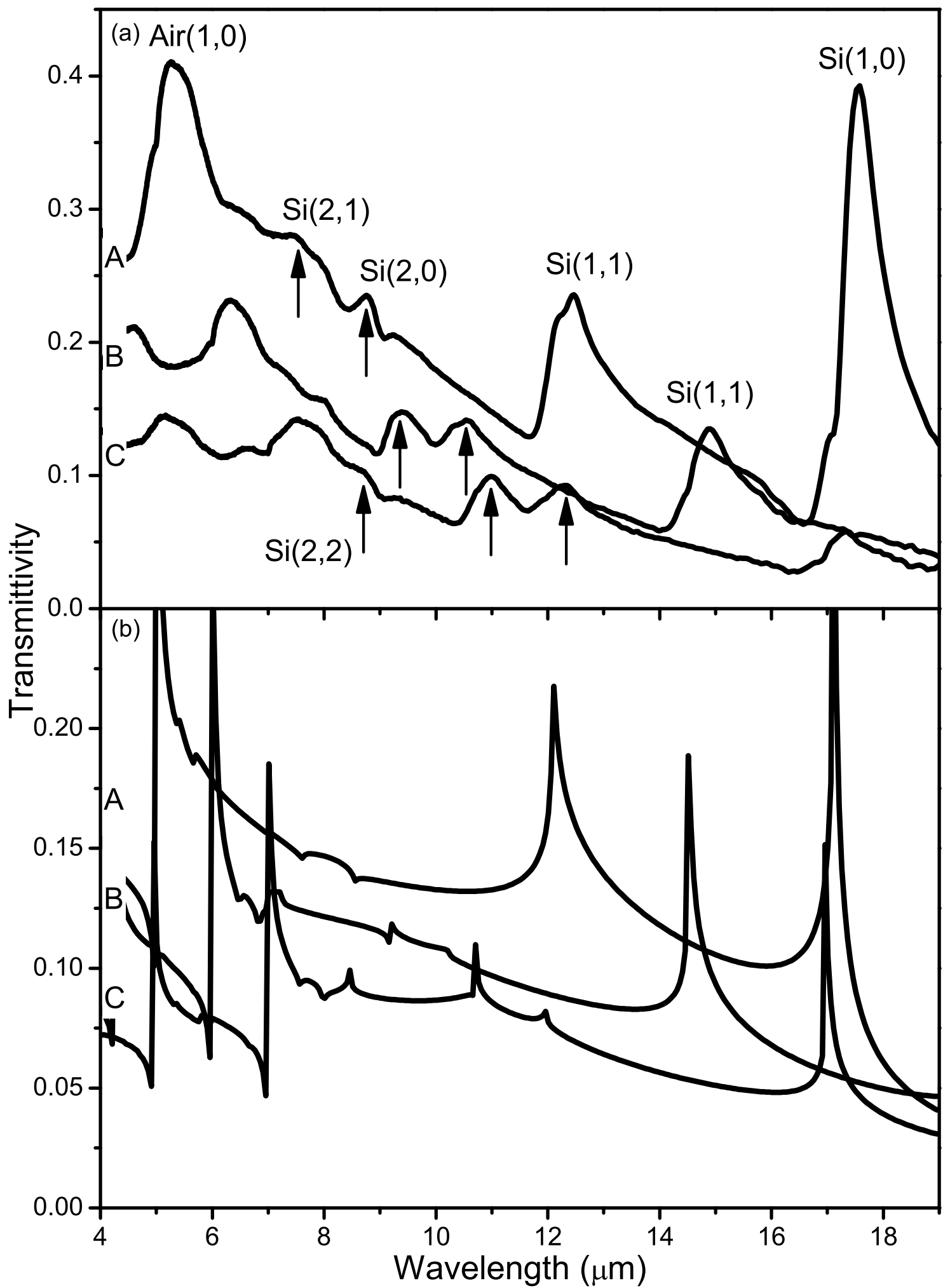
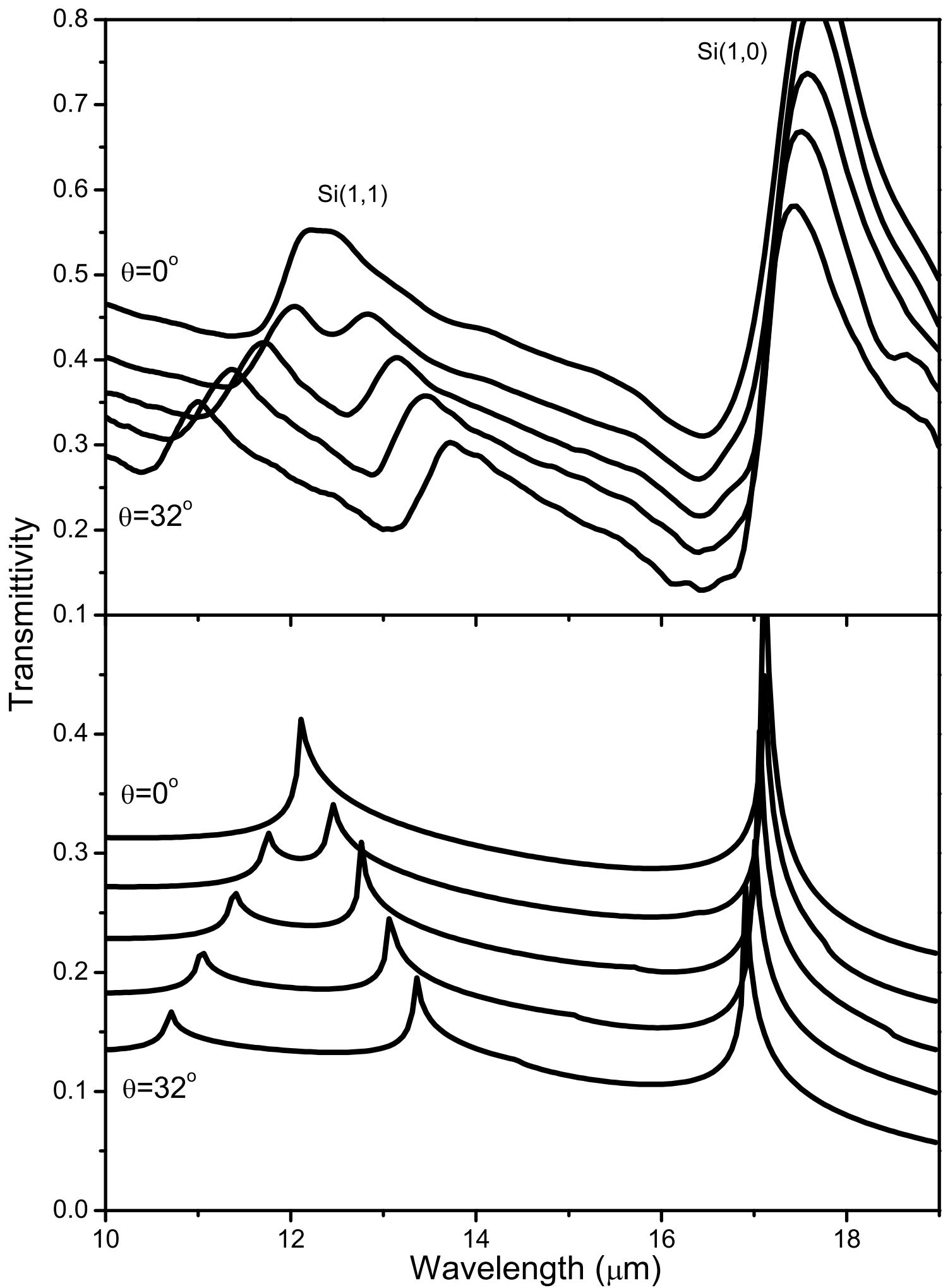
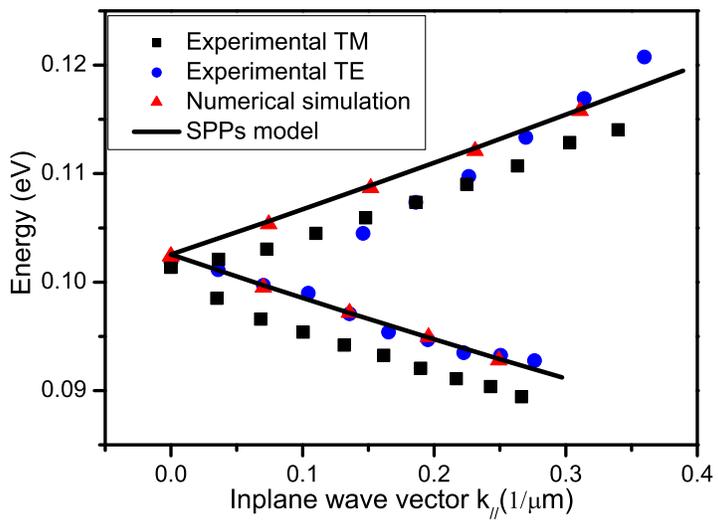
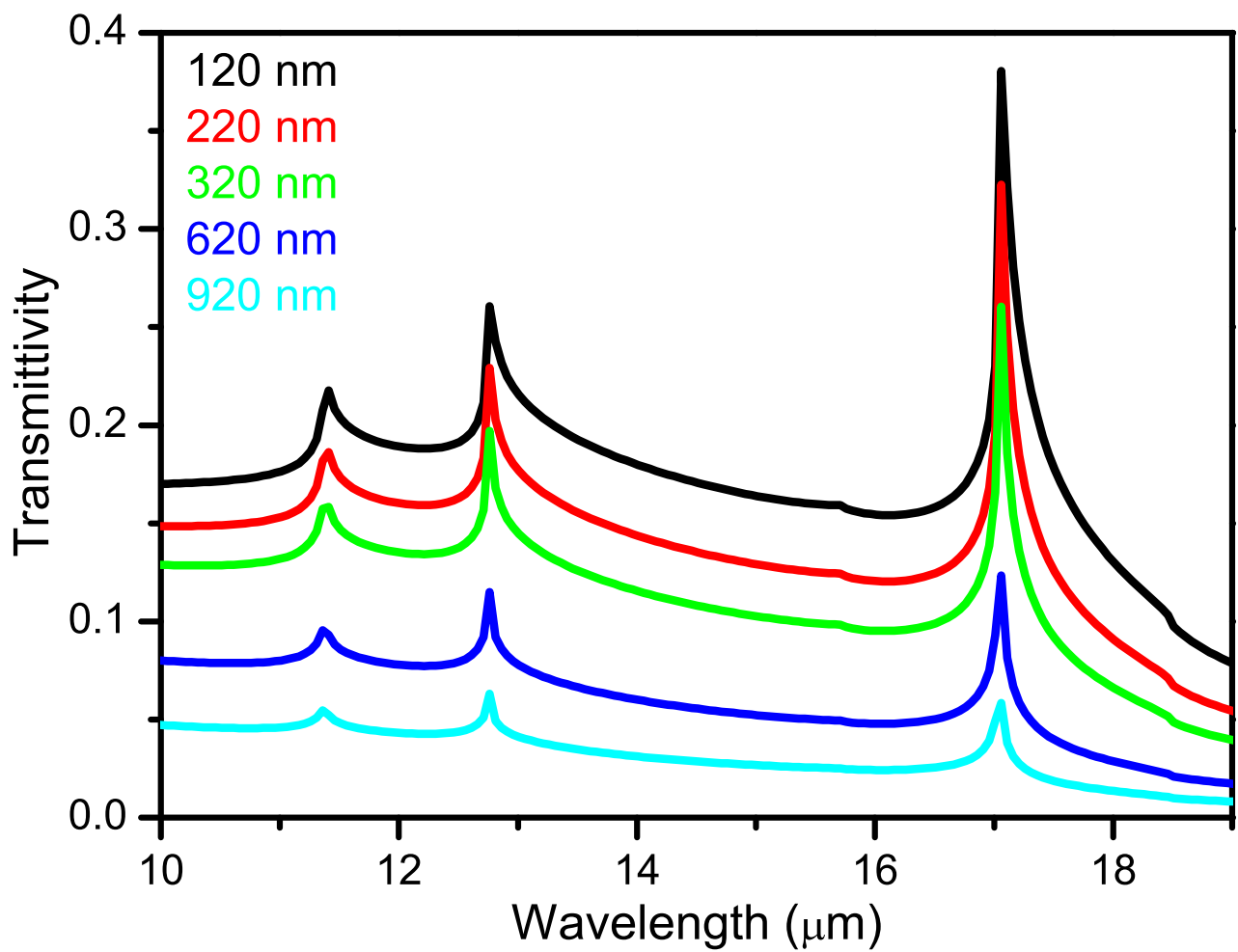
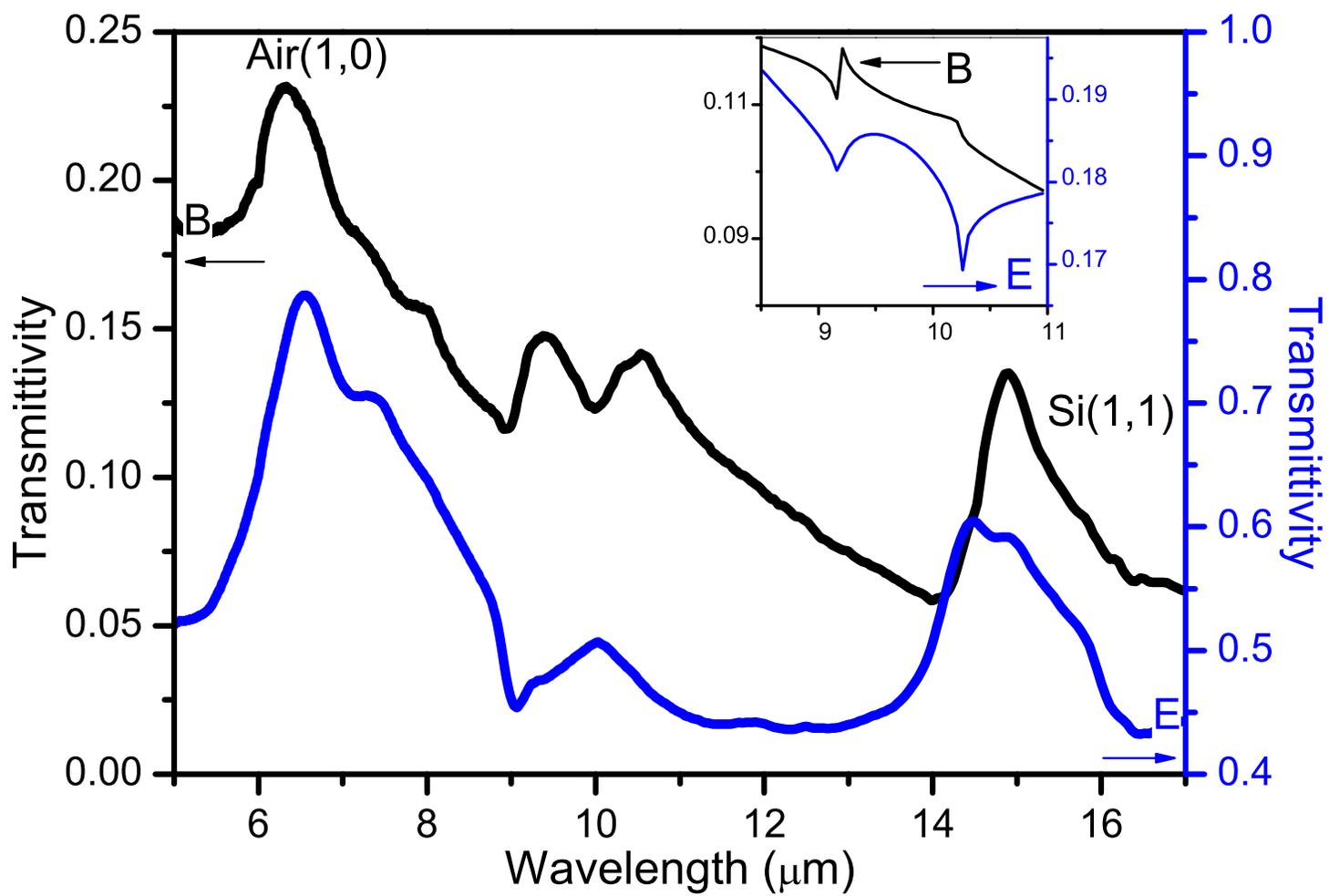
\end{document}